\documentclass{article}

\usepackage{spconf,amsmath,amssymb,graphicx,hyperref}

\usepackage{microtype}
\usepackage{graphicx}
\usepackage{subfigure}
\usepackage{booktabs} % for professional tables
\usepackage{array}

\usepackage{algorithm}
\usepackage{algpseudocode}
\usepackage[subtle]{savetrees}
\ninept
\usepackage{makecell,xcolor}
\usepackage{hyperref}
\usepackage{multirow}
\usepackage[table]{xcolor}
\usepackage{adjustbox}
\usepackage{graphicx}    % for resizebox / scalebox
\usepackage{anyfontsize} % if you want smaller than \tiny
\usepackage{booktabs}    % for professional table rules
\usepackage[table]{xcolor} % for rowcolor
\usepackage{makecell}    % for multirowcell

\definecolor{top1}{rgb}{0.8, 0.5, 0.5}    % Very Light Dark Red
\definecolor{top2}{rgb}{0.95, 0.7, 0.7}   % Lighter Medium Red
\definecolor{top3}{rgb}{0.98, 0.9, 0.9}  % 
\title{ArrayDPS-Refine: Generative Refinement of Discriminative Multi-channel Speech Enhancement}
\name{Zhongweiyang Xu$^{\star}$\thanks{$^{\star}$Work performed during Z. Xu’s Meta internship.}, Ashutosh Pandey$^{\dagger}$, Juan Azcarreta$^{\dagger}$, Zhaoheng Ni$^{\dagger}$, Sanjeel Parekh$^{\dagger}$, Buye Xu$^{\dagger}$}
\address{$^{\star}$University of Illinois Urbana-Champaign,\quad $^{\dagger}$Reality Labs Research at Meta}
\ninept
\begin{document}

\maketitle
\begin{abstract}
Multi-channel speech enhancement aims to recover clean speech from noisy multi-channel recordings. Most deep learning methods employ discriminative training, which can lead to non-linear distortions from regression-based objectives, especially under challenging environmental noise conditions. Inspired by ArrayDPS for unsupervised multi-channel source separation, we introduce ArrayDPS-Refine, a method designed to enhance the outputs of discriminative models using a clean speech diffusion prior. ArrayDPS-Refine is training-free, generative, and array-agnostic. It first estimates the noise spatial covariance matrix (SCM) from the enhanced speech produced by a discriminative model, then uses this estimated noise SCM for diffusion posterior sampling. This approach allows direct refinement of any discriminative model’s output without retraining. Our results show that ArrayDPS-Refine consistently improves the performance of various discriminative models, including state-of-the-art waveform and STFT domain models. Audio demos are provided at \href{https://xzwy.github.io/ArrayDPSRefineDemo/}{https://xzwy.github.io/ArrayDPSRefineDemo/}.
\end{abstract}
\vspace{-10pt}
\section{Introduction}
\vspace{-3pt}
Speech enhancement addresses the challenge of recovering clean speech from noisy reverberant mixture(s). Deep learning has made remarkable progress in single-channel speech enhancement~\cite{zheng2023sixty}, with notable developments in both discriminative and generative modeling approaches. Discriminative models are trained with a regression loss to directly map noisy features to clean speech features. While these models often achieve impressive gains in objective metrics, they frequently introduce distortions that degrade perceptual quality \cite{distortion} and can negatively impact the performance of downstream systems, such as automatic speech recognition (ASR) \cite{rethinkingdistortions}. These distortions are generally attributed to the application of non-linear neural network processing and to the inherently ill-posed problem of speech enhancement, particularly in low SNR conditions where the speech signal is heavily masked by noise \cite{rethinkingdistortions, spear, lavoce}.

In recent years, diffusion-based generative models have become increasingly popular for improving the perceptual quality of single-channel speech enhancement ~\cite{sgmse, cdiff, usee}. While these generative models achieve superior perceptual quality, they still lag behind state-of-the-art (SOTA) discriminative models in terms of objective metrics and ASR performance. To address this gap, one line of research explores using generative models to improve or refine the output of discriminative models~\cite{diffiner}. In a similar vein, StoRM~\cite{storm} leverages the enhanced output of a discriminative model as a warm initialization for a diffusion-based enhancement model, achieving impressive results. Diffiner~\cite{diffiner} proposes using diffusion denoising restoration model (DDRM)~\cite{ddrm} to further enhance any discriminative model. Despite the strong performance of these approaches, their application has so far been limited to single-channel speech enhancement.

Unlike single-channel speech enhancement, multi-channel speech enhancement uses microphone arrays to apply spatial filtering, allowing for more effective separation of speech and noise \cite{consolidated}. Deep learning based discriminative models have achieved great success in exploiting the spatial information, particularly the ones targeting phase enhancement with either waveform  ~\cite{pcm, tadrn, fasnet} or short-time Fourier transform (STFT) domain processing~\cite{uses, uses2}. Most multichannel speech enhancement methods are designed for fixed microphone arrays, limiting their use across different devices and necessitating device-specific model development. To address this, array-agnostic architectures have been proposed that incorporate processing modules agnostic to the order and number of microphones, eliminating the need for model retraining for each array configuration~\cite{fasnet, tadrn, uses2}.

Although discriminative multi-channel speech enhancement methods generally outperform their single-channel counterparts, they still suffer from nonlinear distortions introduced by deep neural networks, much like single-channel approaches~\cite{spear}. To address these distortions, multi-channel systems can integrate DNNs with traditional spatial filters—such as the minimum variance distortionless response (MVDR) beamformer~\cite{consolidated}—which ensures distortionless output. However, this distortionless property often comes at the expense of increased residual noise, necessitating additional post-processing~\cite{consolidated}.

In response to these challenges, there is a growing trend toward generative modeling-based speech enhancement~\cite{mcdiff, mcdiff2, mcdiff3}. 
Recently, ArrayDPS~\cite{arraydps} has demonstrated unsupervised, array-agnostic, and generative multi-channel speech separation by leveraging a pre-trained speech diffusion prior. Not only does this approach achieve performance on par with SOTA discriminative models, but it also introduces a generic diffusion posterior sampling method~\cite{dps} that can be applied to a wide range of multi-channel blind inverse problems~\cite{usddps}. 

Building on these advances, we introduce ArrayDPS-Refine—a training-free, generative, and array-agnostic refinement method designed to enhance any discriminative multi-channel speech enhancement model. For a given discriminative model, ArrayDPS-Refine first estimates the noise spatial covariance matrix (SCM) using the noisy mixture and the enhanced speech from model. The estimated noise SCM is then used to provide likelihood guidance within a modified ArrayDPS framework.

%Despite their promise in mitigating distortions, generative approaches have yet to become mainstream, as they currently lag behind SOTA discriminative multi-channel models in overall performance~\cite{mcdiff3}.
% ~\cite{diffiner} uses a diffusion refiner to further enhance the beamformer output, but the refinement is single-channel which does not fully exploit the multi-channel noisy input
% To solve the distortion problem, some diffusion-based multi-channel speech enhancement models have been proposed~\cite{mcdiff}, but none have shown competitive result compared to SOTA discriminative multi-channel models.

Extensive evaluations show that ArrayDPS-Refine delivers significant improvements in intelligibility, quality, and word error rate (WER) metrics across a range of discriminative models, underscoring its effectiveness and versatility. Furthermore, the observed gains in WER, along with demo samples, demonstrate that ArrayDPS-Refine effectively reduces distortions introduced by discriminative models.

ArrayDPS-Refine's main contributions are: (1) a \textbf{training-free}, \textbf{generative}, \textbf{array-agnostic} method to refine \textbf{any} discriminative enhancement model without any re-training (2) a novel approach of utilizing noise spatial covariance matrix for likelihood guidance in diffusion posterior sampling (3) the first multi-channel generative method that can outperform SOTA discriminative models in \textbf{perceptual}, \textbf{intelligibility} and \textbf{WER} metrics.

\vspace{-10pt}
\section{Background and Problem Formulation}
\vspace{-3pt}
In noisy reverberant scenarios, with a single target speaker and $C$ microphones, let $X(\ell,k)\in\mathbb{C}$ denote the clean anechoic speech at frame $\ell\in[0, L-1]$ and frequency $k\in[0, K-1]$ in the short-time Fourier transform (STFT) domain. The $C$-channel noisy observations in the STFT domain are modeled as
\begin{equation}\label{eq:signal_model}
Y(\ell,k) = H(\ell,k) *_\ell X(\ell,k) + N(\ell,k),
\end{equation}
where $Y(\ell,k)\in\mathbb{C}^C$ denotes the vector of observed multi-channel noisy signals, $N(\ell,k)\in\mathbb{C}^C$ denotes additive noise, $H(\ell,k)\in\mathbb{C}^C$ represents the room acoustic transfer functions (ATF) from the source to the $C$ microphones, and $*_\ell$ denotes convolution across frames. Note that $H\in\mathbb{C}^{N_H\times K\times C}$ is a multi-frame filter with frame length $N_H$.
% The spatial covariance of the noise is characterized by the spatial covariance matrix (SCM)
% \begin{equation}
% \Phi_{NN}(\ell,k) = \mathbb{E}\big[\, N(\ell,k) N(\ell,k)^\mathsf{H} \,\big],
% \end{equation}
% where $(\cdot)^\mathsf{H}$ denotes the Hermitian transpose. 

In this paper, we assume $N(\ell,k)$ follows a zero-mean complex Gaussian distribution with a spatial covariance $\Phi_{\text{NN}}(\ell,k) = \mathbb{E}\big[\, N(\ell,k) N(\ell,k)^{H} \,\big]$, the likelihood of the observations is then:
{
\footnotesize
\begin{equation}\label{eq:likelihood}
p\big(Y(\ell,k)\,|\,H(\ell,k), s(\ell,k)\big) 
= \mathcal{CN}\!\Big(H(\ell,k) *_\ell X(\ell,k), \Phi_{\text{NN}}(\ell,k)\Big)
\end{equation}
}
Thus, the log-likelihood can be calculated analytically with $\Phi_{\text{NN}}$:
{
\footnotesize
\begin{equation}
\mathcal{L} = \sum_{\ell,k} \log p\big(Y(\ell,k)\,|\,H(\ell,k), X(\ell,k)\big)
\end{equation}
}
% Assuming $N(\ell,k)$ follows a zero-mean complex Gaussian distribution with covariance $\Phi_{NN}(\ell,k)$, the conditional distribution of the observations is given by
% \[
% p\big(Y(\ell,k)\,|\,H(\ell,k) *_\ell s(\ell,k)\big) = \mathcal{N}_{\mathbb{C}}\!\Big(H(\ell,k) *_\ell s(\ell,k), \Phi_{NN}(\ell,k)\Big),
% \].

\vspace{-13pt}
\subsection{Diffusion Model}\label{sec:diffusion}
In this paper, we follow Denoising Diffusion Probabilistic Model (DDPM)~\cite{ddpm, improved_ddpm}. Given a data distribution $p_{\text{data}}(x_0)$, DDPM defines a forward diffusion process to gradually transform $x_0$ to $x_1, x_2, ..., x_T$, where $T$ is the final diffusion step, and $x_T\sim\mathcal{N}(0,I)$, following:
\vspace{-3pt}
\begin{align}
q(x_t \!\mid x_{t-1})
  &= \mathcal{N}\!\big(\sqrt{\alpha_t}\,x_{t-1},\,\beta_t I\big)
% q(x_t \!\mid x_0)
%   &= \mathcal{N}\!\big(\sqrt{\bar\alpha_t}\,x_0,\,(1-\bar\alpha_t) I\big),
\end{align}
where $\{\beta_t\}_{t=1}^T$ is a pre-defined noise variance schedule with $\beta_t\in(0,1)$. Then $\alpha_t:=1-\beta_t$ and
$\bar\alpha_t:=\prod_{s=1}^t \alpha_s$, which infers that
\begin{equation}
x_t = \sqrt{\bar\alpha_t}\,x_0
      + \sqrt{1-\bar\alpha_t}\,\epsilon,\qquad \epsilon\sim\mathcal{N}(0,I).
\end{equation}
DDPM proposes that we can train a model to reverse the diffusion process from $x_T~\sim~\mathcal{N}(0,I)$ to $x_0$ step by step, which enables sampling from the data distribution $p_{\text{data}}(x_0)$. Each reversal step is approximated by sampling from $p_\theta(x_{t-1}\!\mid x_t)=\mathcal{N}\!\big(\mu_\theta(x_t,t),\,\sigma^2_t I\big)$, where $\sigma^2_t$ can be derived to be $\sigma^2_t=\frac{1-\Bar{\alpha}_{t-1}}{1-\Bar{\alpha}_{t}}\beta_t$, and $\mu_\theta(x_t, t)$ is:
\vspace{-3pt}
\begin{align}\label{eq:mu}
\mu_\theta(x_t,t) = \frac{1}{\sqrt{\alpha_t}}
   \!\left(
     x_t - \frac{\beta_t}{\sqrt{1-\bar\alpha_t}}\,
     \epsilon_\theta(x_t,t)
   \right),
\end{align}
where a neural network $\epsilon_\theta$ is trained to predict the forward noise by minimizing $\mathbb{E}_{t,x_0\sim~p_{\text{data}},\epsilon\sim\mathcal{N}(0,I)}\!
\Big[
\big\|
\epsilon - \epsilon_\theta(
\sqrt{\bar\alpha_t}\,x_0 + \sqrt{1-\bar\alpha_t}\,\epsilon,\; t)
\big\|_2^2
\Big]$.
The error estimation function $\epsilon_\theta(x_t, t)$ can also be used to denoise $x_t$ to $x_0$ as an MMSE denoiser:
% \vspace{-3pt}
\begin{equation}\label{eq:mmse}
\setlength{\abovedisplayskip}{3pt} % Reduce space above the equation
\setlength{\belowdisplayskip}{3pt} % Reduce space below the equation
    \mathbb{E}[x_0 \!\mid x_t] \simeq \hat{x}_0(x_t,t) = \frac{x_t - \sqrt{1-\bar\alpha_t}\,\epsilon_\theta(x_t,t)}{\sqrt{\bar\alpha_t}}
\end{equation}

Fundamentally equivalent to DDPM, score-based diffusion~\cite{score} defines the forward diffusion process as a stochastic differential equation (SDE), and then uses a neural network $s_\theta(x_t, t)$ to model the score function $\nabla_{x_t}\log p_t(x_t)$, which is used in a derived reversal SDE for sampling. From the Tweedie's formula $\mathbb{E}[x_0 \!\mid x_t] = \frac{1}{\sqrt{\bar\alpha_t}}
   x_t + (1-\bar\alpha_t)\, \nabla_{x_t}~\log p_t(x_t)$ and Eq.~\ref{eq:mmse}, we can also approximate the score function in DDPM by:
\begin{equation}
\setlength{\abovedisplayskip}{3pt} % Reduce space above the equation
\setlength{\belowdisplayskip}{3pt} % Reduce space below the equation
    \nabla_{x_t}\log \;p_t(x_t)\simeq~s_\theta(x_t, t)=-\frac{1}{\sqrt{1-\Bar{\alpha}_t}}\epsilon_\theta(x_t, t)
\end{equation}
% This also allows rewriting Eq.~\ref{eq:mu} to be:
% \begin{align}\label{eq:mu2}
% \mu_\theta(x_t,t) = \frac{1}{\sqrt{\alpha_t}}
%    \!\left(
%      x_t + \beta_t\,s_\theta(x_t, t)
%    \right),
% \end{align}
% \begin{align}
% \mathcal{L}_{\text{simple}}
% = \mathbb{E}_{t,x_0,\epsilon}\!
% \Big[
% \big\|
% \epsilon - \epsilon_\theta(
% \sqrt{\bar\alpha_t}\,x_0 + \sqrt{1-\bar\alpha_t}\,\epsilon,\; t)
% \big\|_2^2
% \Big].
% \end{align}

\vspace{-10pt}
\subsection{Diffusion Posterior Sampling and ArrayDPS}
\vspace{-3pt}
Following Eq.~\ref{eq:signal_model}, where $Y = H *_{\ell} X + N$ for abbreviation, assume the room acoustic transfer function $H$ is known and $N\sim \mathcal{CN}(0, \Phi_{\text{NN}})$. To recover $X$ from measured $Y$, diffusion posterior sampling (DPS)~\cite{dps} proposes to sample from $P(X|Y)$ using a pre-trained score-based diffusion model for clean speech $X$. Following Sec.~\ref{sec:diffusion}, $s_\theta(X_t, t)$ is trained to approximate $\nabla_{X_t}~\log~p(X_t)$, which can be directly used to sample from $P_\text{data}(X)$ following the reverse diffusion process.

To sample from $p(X|Y)$, $\nabla_{X_t}\log p(X_t|Y)$ is needed for the reverse diffusion process, so DPS decomposes $\nabla_{X_t}\log~p(X_t|Y)$ using Bayes' theorem:
\begin{equation}\label{eq:bayes}
\setlength{\abovedisplayskip}{3pt} % Reduce space above the equation
\setlength{\belowdisplayskip}{3pt} % Reduce space below the equation
    \nabla_{X_t}~\log~p(X_t|Y) = \nabla_{X_t}~\log~p(X_t) + \nabla_{X_t}~\log~p(Y|X_t)
\end{equation}
In Eq.~\ref{eq:bayes}, $\nabla_{X_t}\log~p(X_t)$ is approximated by the pre-trained diffusion score model $s_\theta(X_t, t)$, and DPS proposes to approximate the likelihood score $\nabla_{X_t}~\log~p(Y|X_t)$ using:
\begin{align}
\setlength{\abovedisplayskip}{3pt} % Reduce space above the equation
\setlength{\belowdisplayskip}{3pt} % Reduce space below the equation
    \nabla_{X_t}~\log~p(Y|X_t) &\simeq~\nabla_{X_t}~\log~p(Y|\hat{X}_0(X_t, t), H)\label{eq:likelihood_score}\\
    \hat{X}_0(X_t, t) &= \frac{1}{\sqrt{\bar\alpha_t}}
   X_t + (1-\bar\alpha_t)\, s_\theta(X_t, t)\label{eq:tweedie2}
\end{align}
Eq.~\ref{eq:tweedie2} uses Tweedie's formula as mentioned in Sec.~\ref{sec:diffusion}, which denoises $X_t$ to $\hat{X}_0$. Then the denoised $\hat{X}_0$ is used to calculate the likelihood $p(Y|\hat{X}_0, H)$, which can be directly calculated using Eq.~\ref{eq:likelihood} because $H$ is assumed to be known in DPS. However, in real-world scenarios, the room acoustic transfer function $H$ is never known. In the next section we will discuss how ArrayDPS solves this problem.

\vspace{-10pt}
\subsection{ArrayDPS and FCP}
\vspace{-3pt}
ArrayDPS~\cite{arraydps} proposes to use diffusion posterior sampling for multi-channel speech separation under weak white noise. Thus our problem formulation is the same as ArrayDPS with number of speakers set to one under Gaussian noise assumption. To solve the problem of unknown acoustic room transfer function, ArrayDPS proposes to use Forward Convolutive Prediction (FCP)~\cite{fcp} to estimate $H$ for each speech source at each DPS step where $H$ is needed for likelihood calculation. 

Intuitively, forward convolutive prediction (FCP)~\cite{fcp} is an STFT-domain filter estimation algorithm that tries to find the best filter such that the filtered input matches the target the most. To estimate the room acoustic transfer functions, a source signal $X(\ell,k)\in\mathbb{C}$ is input to FCP, and the $C$-channel noisy mixture $Y(\ell,k)\in\mathbb{C}^C$ is the FCP target. We denote $Y^c\in\mathbb{C}(\ell,k)$ as $c^{\text{th}}$ channel noisy mixture and $H^c(\ell,k)$ as $c^{\text{th}}$ channel acoustic transfer function. Then FCP($X$, $Y$) estimates each channel's room acoustic transfer functions by solving:

% Given any input $X(l,k)\in\mathbb{C}$ and target $Y(l,k)\in\mathbb{C}$ (X and Y here are only for illustration of FCP, please do not confuse with the noisy and clean signals in Eq.~\ref{eq:signal_model}). Then FCP 
{\footnotesize
\vspace{-5pt}
\setlength{\abovedisplayskip}{0pt} % Reduce space above the equation
\setlength{\belowdisplayskip}{2pt} % Reduce space below the equation
% \vspace{-3pt}
\begin{align}
    &\hat{H}^c = \underset{{H}^c}{\arg\min} \sum_{m,k} \frac{1}{\hat{\lambda}_{m,k}} \left|Y^c({m,k}) - \sum_{n=0}^{N_H-1} H^c({n,k}) {X}({m-n,k})\right|^2\label{eq:fcp_obj}\\
    &\hat{\lambda}_{m,k} = \frac{1}{C} \sum_{c=1}^C |Y^{c}({m, k})|^2+ \epsilon\cdot \max\limits_{m,k}\frac{1}{C} \sum_{c=1}^C |Y^{c}({m,k})|^2\label{eq:fcp_lambda}
\end{align}
}
As shown above, FCP is solving a weighted least squares problem, so it has an analytical solution as in~\cite{fcp}. $N_H$ is the number of frames of the acoustic transfer function. The weight $\hat{\lambda}_{m,k}$ aims to prevent the estimated filters from overfitting to high-energy STFT bins, and $\epsilon$ is a tunable hyperparameter to adjust the weight.
\begin{figure*}[t]
    \centering
    \vspace{-45pt}
    \includegraphics[width=0.9\linewidth]{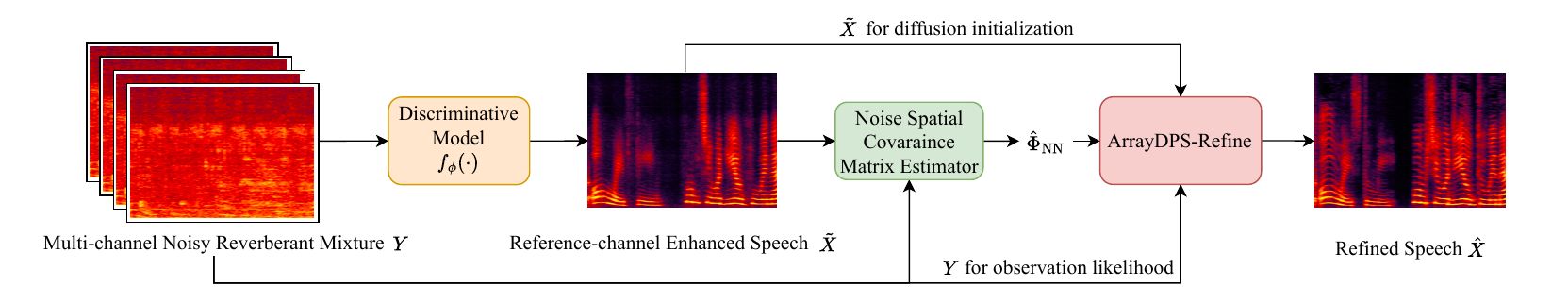}
    \vspace{-15pt}
    \caption{ArrayDPS-Refine Pipeline. }
    \vspace{-10pt}
    \label{fig:refinement_pipeline}
\end{figure*}
\begin{algorithm}[t]
\caption{ArrayDPS-Refine}
\label{alg:inference}
\begin{algorithmic}[1]
% \Require multi-channel noisy reverberant speech ${Y}$
\Require $Y, \hat{\Phi}_{\text{NN}}, \tilde{X}, T', \xi, \epsilon_\theta(\cdot), \{\sigma^2_t\}_{t=1}^{T'}, \{\alpha_t\}_{t=1}^{T'}, \{\bar{\alpha}_t\}_{t=1}^{T'}$
\State $\tilde{X}' \gets |\tilde{X}|^{0.5}\,\exp\!\big(j\,\angle \tilde{X}\big)$ \Comment{\textcolor{magenta}{transform to compressive domain}}
\State Sample $\epsilon\sim\mathcal{CN}(0,2I)$ \Comment{\textcolor{magenta}{$\mathcal{N}(0, I)$ for both real and imag}}
\State $X_{T'}'\gets \sqrt{\bar\alpha_{T'}}\,\tilde{X}'
      + \sqrt{1-\bar\alpha_{T'}}\,\epsilon$ \Comment{\textcolor{magenta}{init. to step T'}}
\For{$t = T', \dots, 1$}
    \State $\hat{\epsilon} \gets \epsilon_\theta(X_t', t)$ \Comment{\textcolor{magenta}{diffusion model estimate noise}}
    \State Sample $Z\sim\mathcal{CN}(0,2I)$ \Comment{\textcolor{magenta}{$\mathcal{N}(0, I)$ for both real and imag}}
    \State $X_{t-1}'\gets \frac{1}{\sqrt{\alpha_t}}(X_t'-\frac{1-\alpha_t}{\sqrt{1-\bar{\alpha}_t}}{\hat{\epsilon}})+\sigma^2_tZ$ \Comment{\textcolor{magenta}{prior step}}
    \State $\hat{X}_0'\gets \frac{1}{\sqrt{\bar{\alpha}_t}}(X_t'-\sqrt{1-\bar{\alpha}_t}\hat{\epsilon})$ \Comment{\textcolor{magenta}{one-step denoising}}
    \State $\hat{X}_0 \gets |\hat{X}_0'|^2\,\exp\!\big(j\,\angle \hat{X}_0'\big)$ \Comment{\textcolor{magenta}{transform to STFT domain}}
    \State $\hat{H}\gets\text{FCP}(\hat{X}_0, Y)$ \Comment{\textcolor{magenta}{room ATF estimation}}
    \State $\hat{X}_\text{reverb}\gets \hat{H}*_\ell \hat{X}_0$ \Comment{\textcolor{magenta}{estimate multi-channel reverb. speech}}
    \State $\hat{N} = Y-\hat{X}_{\text{reverb}}$\Comment{\textcolor{magenta}{estimate multi-channel noise}}
    % \For{$l = 0, \dots, L-1$, $k = 0, \dots, K-1$}
    % \State $X_{t-1}'\gets X_{t-1}' - \xi\frac{1-\alpha_t}{\sqrt{\alpha_t}}\nabla_{X_t'}\sum_{l,k}\frac{1}{2}\hat{N}(l,k)^H\hat{\Phi}^{-1}_{\text{NN}}(l,k)\hat{N}(\ell,k)$
{\footnotesize    \State $G\gets\nabla_{X_t'} \!\left[
      -\tfrac{1}{2}\sum\limits_{\ell,k}
      \hat{N}(\ell,k)^{\mathrm{H}}\,
      \widehat{\Phi}_{\mathrm{NN}}^{-1}(\ell,k)\,
      \hat{N}(\ell,k)
    \right]$\Comment{\textcolor{magenta}{likelihood score}}}
\State $\displaystyle
X_{t-1}' \gets X_{t-1}' + \xi\,\frac{1-\alpha_t}{\sqrt{\alpha_t}}\,
G$ \Comment{\textcolor{magenta}{likelihood step}}
\EndFor
\State ${X}_0 \gets |{X}_0'|^2\,\exp\!\big(j\,\angle {X}_0'\big)$ \Comment{\textcolor{magenta}{transform to STFT domain}}
\State $H_{\text{align}}\gets \text{FCP}(X_0, \tilde{X})$ \Comment{\textcolor{magenta}{single-frame filter est. for alignment}}
\State \Return $H_{\text{align}}*_\ell X_0$ \Comment{\textcolor{magenta}{return aligned signal}}
\end{algorithmic}
\end{algorithm}
Thus, similar to Eq.~\ref{eq:arraydps_likelihood_score}, at each DPS step, ArrayDPS approximates the likelihood score $\nabla_{X_t}~\log~p(Y|X_t)$ with:
% \vspace{-3pt}
{\footnotesize
\begin{align}
\setlength{\abovedisplayskip}{2pt} % Reduce space above the equation
\setlength{\belowdisplayskip}{2pt} 
&\nabla_{X_t}~\log~p(Y|X_t)~\simeq~\nabla_{X_t}~\log~p(Y|\hat{X}_0(X_t, t), \hat{H})\label{eq:arraydps_likelihood_score}\\
&\hat{X}_0(X_t, t)=\frac{1}{\sqrt{\bar\alpha_t}} X_t + (1-\bar\alpha_t)\, s_\theta(X_t, t),\;\;\hat{H} = \text{FCP} (\hat{X}_0(X_t, t), Y)\label{eq:arraydps_fcp}
\end{align}}
Eq.~\ref{eq:arraydps_fcp} first denoises $X_t$ to $\hat{X}_0$, and uses $\hat{X}_0$ to estimate $\hat{H}$ using FCP. Then $\hat{H}$ is used for likelihood calculation in Eq.~\ref{eq:arraydps_likelihood_score}.

\vspace{-10pt}
\section{Method}
\vspace{-3pt}
We propose an ArrayDPS-based refiner for multi-channel speech enhancement. The overall pipeline is shown in Fig.~\ref{fig:refinement_pipeline}. A discriminative enhancement model $f_\phi(\cdot)$ first processes a multi-channel mixture $Y \in \mathbb{C}^{L \times K \times C}$ (with $L$ time frames and $K$ STFT frequency bins), yielding 
$\tilde{X} = f_\phi(Y) \in \mathbb{C}^{L \times K}$,
the reference-channel anechoic clean speech estimate. Then $\tilde{X}$ and $Y$ are used to estimate the noise spatial covariance matrix (SCM) $\Phi_{\text{NN}}(\ell,k)\in\mathbb{C}^{C\times C}, l\in[0, L-1], k\in[0, K-1]$. With the noise SCM estimated, the likelihood as in Eq.~\ref{eq:likelihood} can be calculated, which allows us to refine $\tilde{X}$ using ArrayDPS-Refine in Algorithm~\ref{alg:inference}.

\vspace{-10pt}
\subsection{Spatial Covariance Matrix Estimation}
With the reference-channel clean speech estimate $\tilde{X}$ estimated by a discriminative model, we first estimate the room acoustic transfer function using FCP: $\tilde{H}=\text{FCP}(\tilde{X}, Y)$. This allows us to estimate all-channel reverberant clean speech: $\tilde{X}_{\text{reverb}}=\tilde{H}*_\ell \tilde{X}$, where $\tilde{X}_{\text{reverb}}\in\mathbb{C}^{L\times K \times C}$. Then the multi-channel noise can be estimated by $\tilde{N}=Y-\tilde{X}_{\text{reverb}}$, which can then be used to estimate the noise SCM $\Phi_{\text{NN}}$:
\begin{align}\label{eq:ema}
\setlength{\abovedisplayskip}{2pt} % Reduce space above the equation
\setlength{\belowdisplayskip}{2pt} % Reduce space below the equation
    \hat{\Phi}_{\text{NN}}(\ell,k) = \alpha\hat{\Phi}_{\text{NN}}(\ell-1,k) + (1-\alpha) \tilde{N}(\ell,k)\tilde{N}^H(\ell,k)
\end{align}
As in Eq.~\ref{eq:ema}, we estimate the noise SCM using recursive exponential moving average with a smoothing coefficient, $\alpha$. Then the estimated noise SCM can be used for likelihood calculation in ArrayDPS-Refine.

\vspace{-10pt}
\subsection{ArrayDPS-Refine}\label{sec:arraydps_refine}
The ArrayDPS-Refine algorithm is shown in Algorithm~\ref{alg:inference}, which uses a pre-trained diffusion model to refine discriminative model's enhanced speech $\tilde{X}$. Different from ArrayDPS, we train a prior diffusion model on compressive-domain STFT of clean speech. Given any clean speech STFT $X$ from a dataset, the compressive STFT is $X'=|{X}|^{0.5}\,\exp\!\big(j\,\angle {X}\big)$. Previous work has shown the effectiveness of using magnitude compression for audio diffusion models~\cite{edmsound}. 

Following Algorithm~\ref{alg:inference}, ArrayDPS-Refine first transforms $\tilde{X}$ to compressive domain $\tilde{X}'$ for diffusion initialization (line 1). Then in line 2-3, $X'_{T'}$ is initialized from $\tilde{X}'$ to diffusion time step $T'\in[1, T]$ (Similar to StoRM~\cite{storm}, we start from intermediate diffusion step $T'$ initialized from $\tilde{X}'$). In line 2, the noise variance is set to 2 because the diffusion takes place in the real and imaginary component of the compressive STFT, but the algorithm is defined in complex domain. Then line 4-14 are the diffusion posterior sampling process. Line 5-7 use the pre-trained prior diffusion model for one reversal step. Then line 8 gets the one-step denoised compressive STFT $\hat{X}_0'$, which is then transformed to STFT $\hat{X}_0$ in line 9. Same as ArrayDPS estimating $H$ for likelihood estimation, $\hat{X}_0$ is used to estimate the room ATF $H$ in line 10. Using the likelihood model in Eq.~\ref{eq:likelihood} and the estimated noise SCM $\hat{\Phi}_{\text{NN}}$, line 11-13 calculates $G$, the likelihood score approximation, similar to Eq.~\ref{eq:arraydps_likelihood_score} for ArrayDPS. Line 14 then uses $G$ for a likelihood score update, with a scalar $\xi$ controlling the likelihood guidance.

\begin{table*}[t]
\scriptsize
\vspace{-40pt}
\centering
\caption{ArrayDPS-Refine evaluation on Adhoc-microphone arrays (4-channel and 8-channels).}
\setlength\tabcolsep{0.7pt}
\renewcommand{\arraystretch}{0.3}
\label{tab:new_eval}
\begin{tabular}{@{}c c |c | ccccccc | ccccccc@{}}
\toprule
\multirowcell{2}{Row} &
\multirowcell{2}{Methods} &
\multirowcell{2}{$\xi$} &
% \multirowcell{2}{$T'$} &
\multicolumn{7}{c|}{\textbf{4-channel}} &
\multicolumn{7}{c}{\textbf{8-channel}} \\
\cmidrule(lr){4-10} \cmidrule(l){11-17}
& & &
STOI & eSTOI & { PESQ(NB/WB)} & SI-SDR & WER(\%) & DNSMOS & UTMOSv2 &STOI & eSTOI & {PESQ(NB/WB)} & SI-SDR & WER(\%) & {DNSMOS} & UTMOSv2 \\
\midrule
\rowcolor{gray!10}
A0 & Noisy & - &
0.631 & 0.361 & 1.40 / 1.07 & -6.3 & 118 & 1.48 & 1.82 &
0.631 & 0.361 & 1.40 / 1.07 & -6.3 & 118 & 1.48 & 1.82\\
\midrule
A1 & TADRN\cite{tadrn} & - &
0.896 & 0.783 & 2.86 / 2.05 & 8.9 & 58.4 & 2.84 & 2.55 &
0.913 & 0.814 & 2.98 / 2.21 & 9.9 & 46.9 & 2.85 & 2.68 \\
\rowcolor{gray!10}
A2 & TADRN+MCWF & - &
0.785 & 0.562 & 2.01 / 1.23 & 4.6 & 61.9 & 1.55 & 1.74 &
0.845 & 0.655 & 2.17 / 1.32 & 6.8 & 47.0 & 1.66 & 1.88 \\
% 1 & TADRN+MCWF & - & - &
% 0.910 & 0.811 & 2.97 / 2.21 & 9.7 & 47.3 & 2.85 & &
% 0.910 & 0.811 & 2.97 / 2.21 & 9.7 & 47.3 & 2.85 &\\
A3 & Refined TADRN (ours) & 0.4 &
0.910 & 0.812 & 2.99 / \textbf{2.23} & 10.0 & 37.0 & \textbf{2.91} & \textbf{2.98} &
0.928 & 0.844 & 3.12 / 2.43 & 11.1 & 32.0 & \textbf{2.91} & \textbf{3.02}\\
\rowcolor{gray!10}
A4 & Refined TADRN (ours) & 0.6 &
0.913 & 0.818 & \textbf{3.00} / \textbf{2.23} & 10.2 & 38.9 & 2.90 & 2.95 &
0.931 & 0.849 & \textbf{3.14} / \textbf{2.44} & 11.3 & 31.1 & \textbf{2.91} & \textbf{3.02}\\
A5 & Refined TADRN (ours) & 0.8 &
\textbf{0.915} & \textbf{0.820} & 2.99 / \textbf{2.24} & \textbf{10.3} & 37.0 & 2.89 & 2.92 &
\textbf{0.933} & \textbf{0.853} & \textbf{3.14} / \textbf{2.44} & {11.4} & 26.7 & 2.90 & 2.98\\
\rowcolor{gray!10}
A6 & Refined TADRN (ours) & 1.0 &
\textbf{0.915} & \textbf{0.820} & 2.97 / 2.21 & \textbf{10.3} & \textbf{32.8} & 2.88 & 2.88 &
\textbf{0.933} & \textbf{0.853} & 3.13 / \textbf{2.44} & \textbf{11.5} & \textbf{26.1} & 2.88 & 2.94\\
A7 & Refined TADRN (ours) & 1.2 &
\textbf{0.915} & \textbf{0.820} & 2.95 / 2.18 & \textbf{10.3} & 34.9 & 2.86 & 2.83 &
0.930 & 0.840 & 3.10 / 2.41 & {11.3} & 27.8 & 2.86 & 2.87\\
\midrule
\rowcolor{gray!10}
B1 & FASNET-TAC\cite{fasnet} & - &
0.839 & 0.677 & 2.52 / 1.65 & 5.4 & 68.9 & 2.57 & 1.82 &
0.856 & 0.708 & 2.61 / 1.74 & 6.2 & 58.4 & 2.69 & 1.91 \\
B2 & FASNET-TAC+MCWF & - &
0.774 & 0.548 & 2.01 / 1.24 & 3.3 & 66.0 & 1.54 & 1.69 &
0.830 & 0.632 & 2.16 / 1.32 & 5.2 & 53.0 & 1.64 & 1.83 \\
% 1 & FASNET-TAC+MCWF & - & - &
%  &  &  /  &  &  &  & &
%  &  &  /  &  &  &  &\\
\rowcolor{gray!10}
B3 & Refined FASNET-TAC (ours) & 0.4 &
0.867 & 0.736 & \textbf{2.73} / \textbf{1.88} & 6.7 & 56.8 & \textbf{2.77} & \textbf{2.59} &
0.891 & 0.775 & 2.86 / 2.04 & 7.9 & 43.6 & \textbf{2.77} & \textbf{2.68} \\
B4 & Refined FASNET-TAC (ours) & 0.6 &
0.870 & 0.741 & \textbf{2.73} / \textbf{1.88} & 6.8 & 53.3 & 2.73 & 2.53 &
0.894 & 0.781 & \textbf{2.87} / \textbf{2.05} & \textbf{8.0} & 44.4 & 2.75 & 2.61 \\
\rowcolor{gray!10}
B5 & Refined FASNET-TAC (ours) & 0.8 &
\textbf{0.871} & \textbf{0.742} & 2.71 / 1.86 & \textbf{6.9} & 59.3 & 2.71 & 2.48 &
0.895 & \textbf{0.782} & 2.86 / \textbf{2.05} & \textbf{8.0} & \textbf{39.6} & 2.73 & 2.56 \\
B6 & Refined FASNET-TAC (ours) & 1.0 &
\textbf{0.871} & 0.740 & 2.68 / 1.84 & 6.8 & \textbf{47.0} & 2.67 & 2.40 &
\textbf{0.896} & 0.781 & 2.83 / 2.03 & \textbf{8.0} & 41.6 & 2.71 & 2.49 \\
\rowcolor{gray!10}
B7 & Refined FASNET-TAC (ours) & 1.2 &
0.870 & 0.737 & 2.65 / 1.81 & 6.8 & 51.2 & 2.65 & 2.32 &
\textbf{0.896} & 0.780 & 2.81 / 2.01 & \textbf{8.0} & 41.4 & 2.69 & 2.42 \\
% \midrule
% 1 & USES\cite{uses} & - & - &
%  &  &  /  &  &  &  & &
%  &  &  /  &  &  &  &\\
% 1 & USES+MCWF & - & - &
%  &  &  /  &  &  &  & &
%  &  &  /  &  &  &  &\\
% \rowcolor{gray!10}
% 2 & Refined USES & 0.4 & 300 &
%  &  &  /  &  &  &  & &
%  &  &  /  &  &  &  &\\
% 3 & Refined USES & 0.6 & 300 &
%  &  &  /  &  &  &  & &
%  &  &  /  &  &  &  &\\
% \rowcolor{gray!10}
% 4 & Refined USES & 0.8 & 300 &
%  &  &  /  &  &  &  & &
%  &  &  /  &  &  &  &\\

\midrule
C1 & USES2\cite{uses2} & - &
0.923 & 0.833 & \textbf{3.13} / \textbf{2.42} & 6.0 & 36.6 & 2.85 & 2.90 &
0.936 & 0.858 & 3.25 / 2.59 & 5.7 & 30.5 & 2.86 & 3.00\\
\rowcolor{gray!10}
C2 & USES2+MCWF & - &
0.789 & 0.567 & 2.01 / 1.24 & 2.8 & 59.4 & 1.55 & 1.73 &
0.849 & 0.660 & 2.19 / 1.33 & 3.8 & 53.5 & 1.66 & 1.89\\
C3 & Refined USES2 (ours) & 0.4 &
0.922 & 0.835 & {3.11} / {2.41} & 6.7 & {32.7} & 2.88 & \textbf{3.07} &
0.938 & 0.863 & \textbf{3.26} / \textbf{2.63} & 6.2 & 24.2 & \textbf{2.89} & \textbf{3.10}\\
\rowcolor{gray!10}
C4 & Refined USES2 (ours) & 0.6 &
\textbf{0.924} & \textbf{0.837} & 3.10 / 2.40 & \textbf{6.8} & 31.1 & 2.87 & 3.02 &
0.940 & 0.867 & {3.25} / {2.62} & \textbf{6.4} & 24.0 & 2.86 & 3.08\\
C5 & Refined USES2 (ours) & 0.8 &
{0.923} & \textbf{0.838} & 3.07 / 2.37 & \textbf{6.8} & 33.5 & 2.85 & 2.97 &
{0.940} & \textbf{0.868} & 3.23 / 2.60 & \textbf{6.4} & 22.2 & 2.85 & 3.03\\
\rowcolor{gray!10}
C6 & Refined USES2 (ours) & 1.0 &
0.922 & 0.835 & 3.03 / 2.33 & \textbf{6.8} & 29.2 & 2.83 & 2.90 &
\textbf{0.941} & \textbf{0.868} & 3.21 / 2.57 & \textbf{6.4} & \textbf{22.0} & 2.84 & 3.01\\
C7 & Refined USES2 (ours) & 1.2 &
0.920 & 0.831 & 2.98 / 2.26 & 6.7 & \textbf{28.2} & 2.80 & 2.84 &
{0.940} & 0.866 & 3.17 / 2.53 & 6.3 & 22.7 & 2.81 & 2.96\\

\bottomrule
\end{tabular}
\vspace{-10pt}
\end{table*}
With the compressive-domain clean $X_0$ sampled, line 16 transforms it back to STFT domain. However, one problem of this $X_0$ sampled is that there is no constraint in the algorithm making sure that it aligns with the clean signal in the reference channel. To solve this problem, we use another FCP with $N_H=1$ in line 17 to estimate a single-frame filter that can align $X_0$ with $\tilde{X}$. Since the discriminative model is trained to output the reference-channel clean signal, $\tilde{X}$ is fully aligned. Finally, line 18 outputs the aligned version of $X_0$.

% takes the multi-channel reverberant mixture $Y$, estimated noise SCM $\hat{\Phi}_{\text{NN}}$, discriminative enhanced reference-channel speech \tilde{X} as inputs.

\vspace{-10pt}
\section{Experiments and Results}
\vspace{-5pt}
\subsection{ArrayDPS-Refine Configurations}
    As discussed in Sec.~\ref{sec:diffusion}, we follow the DDPM diffusion process~\cite{ddpm}, with the diffusion noise schedule $\beta_t$ increasing linearly from $\beta_1=10^{-4}$ to $\beta_T=0.02$, and $T=1000$ steps. For the diffusion architecture, we use the 2-D U-Net same as in Diffiner~\cite{diffiner}, which accommodates real and imaginary channels of STFT. In our case, the model input is the real and imaginary channels of the compressive STFT as mentioned in Sec.~\ref{sec:arraydps_refine}. We use FFT size of 512, hop size of 128, and square root Hanning window. For diffusion model training, we train on 4-second 16kHz clean speech segments (waveform normalized to $-1$ to $1$) from about 220 hours of clean speech from DNS-Challenge~\cite{dns}, using the Adam optimizer with learning rate $10^{-4}$ and batch size 64. The model is trained for $2.5\times 10^6$ steps on 8 H100 GPUs. For inferece, we use the exponential moving average (EMA) of the model weight with a decay of $0.9999$.

For noise SCM estimation, we set the smoothing factor $\alpha$ in Eq.~\ref{eq:ema} to be $0.95$. For ArrayDPS-Refine in Algorithm.~\ref{alg:inference} and Sec.~\ref{sec:arraydps_refine}, we set the starting diffusion step $T'$ to be $300$. The likelihood score guidance $\xi$ controls the trade-off between better quality (higher prior) and better mixture consistency (higher likelihood). Thus we evaluate result on $\xi\in\{0.4,  0.6, 0.8, 1.0, 1.2\}$. For the FCP in line 10 in Algorithm.~\ref{alg:inference}, we set the filter frame length $N_H$ to be $13$ and $\epsilon$ to be $10^{-3}$, following the configuration in ArrayDPS~\cite{arraydps}. For the alignment FCP in line 17 of Algorithm.~\ref{alg:inference}, $N_H$ is set to $1$ for single frame alignment.

\vspace{-10pt}
\subsection{Discriminative Baselines and Dataset}
\vspace{-3pt}
As we focus on array-agnostic settings, we use three array-agnostic discriminative models as baseline models which our method refines from: FaSNet-TAC~\cite{fasnet}, TADRN~\cite{tadrn}, and USES2~\cite{uses2}. FaSNet-TAC is a time-domain model using transform-and-concatenate (TAC) for array-agnostic multi-channel modeling~\cite{fasnet}. We use the official implementation and configuration\footnote{https://github.com/yluo42/TAC/blob/master/FaSNet.py}. TADRN is a state-of-the-art time-domain enhancement model, which uses a triple-path network for frame-wise, chunk-wise, and channel-wise modeling. We follow the same MIMO configuration as in~\cite{tadrn}. USES2~\cite{uses2} is an STFT-domain SOTA array-agnostic model. We also follow the official implementation and configuration\footnote{https://github.com/espnet/espnet}(USES2-Comp as in~\cite{uses2}), with FFT and hop size to be 512 and 256 respectively, and square root hanning window. For each enhancement model, we also use the enhanced speech to estimate a low-distortion single-frame time-invariant multi-channel Wiener filter (MCWF)~\cite{mcwf} as a baseline. The MCWF is using FFT size of 256 and hop size of 64 respectively.

To train these array-agnostic discriminative models, we create ad-hoc microphone array datasets. For each sample in the dataset, we randomly draw a shoe-box room whose three dimensions range from $3\times3\times2$ to $10\times10\times5$ m. The absorption coefficient is uniformly sampled from $0.3$ to $0.7$ ($T60\in[0.13, 0.55]$ s). The positions of 8 microphones are randomly sampled inside a sphere with radius $0.1 m$, which forms an 8-channel ad-hoc microphone array. The number of interference speakers are randomly sampled from 8-16, simulating bubble noise. The number of noise sources are sampled from 1-50, simulating diverse spatial noise field. All sources' and microphone array center's locations, are all randomly sampled inside the room. The signal-to-noise ratio and signal-to-interference ratio are set to $[-10, 5]$ dB and $[5, 10]$ dB respectively. All the speech sources and noise sources are sampled from DNS-Challenge dataset~\cite{dns}. We use Pyroomacoustics~\cite{pyroomacoustics}'s image method~\cite{imagesource} (order 6) to simulate the noisy mixtures. We simulate $80,000$ $10$-second samples for training, $1,000$ $4$-second samples for validation, $1,000$ $4$-second samples for testing.

During training, the number of channels in each batch is randomly sampled from $2$ to $8$, allowing training on variable number of channels. Also, the models are trained on randomly chunked 4-second segments. A phase constrained magnitude (PCM) loss~\cite{tadrn} is used, with the training target to be the anechoic clean speech. We use the Adam optimizer with a learning rate of $10^{-4}$. For FaSNet-TAC and TADRN, we train the model for 80 epochs with a batch size of 16. For USES2, we train the model for 40 epochs with a batch size of 8.

\vspace{-10pt}
\subsection{Results and Analysis}
\vspace{-3pt}
We evaluate all the discriminative baselines and ArrayDPS-Refine on both 4-channel and 8-channel of the test dataset. For evaluation metrics, we use STOI~\cite{stoi}, eSTOI~\cite{estoi}, and word error rate (WER) to evaluate speech intelligibilty. For WER, we use a Whisper~\cite{whisper} base model for speech recognition. We use SI-SDR~\cite{sisdr} to evaluate sample-level consistency. We use PESQ~\cite{pesq}, DNSMOS~\cite{dnsmos} (overall score), and UTMOSv2~\cite{utmosv2} to evaluate speech perceptual quality. Note that both DNSMOS and UTMOSv2 are non-intrusive. The result is shown in Table.~\ref{tab:new_eval}.

For TADRN's result in Row A1-A7, after refinement, ArrayDPS-refine greatly outperforms TADRN in all metrics. Specifically for Row A3, Refined TADRN improves TADRN by about 0.03 in eSTOI, 0.2 in Wide-band PESQ, 1 dB in SI-SDR, and 0.4 in UTMOSv2 in both 4-channel and 8-channel cases. Also, Refined TADRN decrease the WER by more than 10 percent in 4-channel case, and about 15 percent in 8-channel case, showing remarkable improvement in speech intelligibility. All metrics are improved greatly after refinement.
% showing ArrayDPS-Refine's effectiveness considering TADRN is a SOTA time-domain enhancement model.

For FaSNet-TAC's result in Row B1-B7, the improvement of refinement is even larger. For Row B3 with $\xi=0.4$, Refine-FaSNet-TAC improves FaSNet-TAC with more than 0.06 in eSTOI, 0.2 in wide-band PESQ, 1.3 dB in SISDR, 12 percent in WER, and 0.7 in UTMOSv2, in both 4-channel and 8-channel cases.
% This shows ArrayDPS-Refine's robustness in the preceding discriminative model.

For USES2's result in Row C1-C7, USES2 achieves superior enhancement performance even comparing TADRN (better than TADRN in all metrics except SISDR). For Row C3 with $\xi=0.4$, ArrayDPS-Refine can still consistently improve over USES2. In 4-channel case, Refined USES2 achieves similar performance to USES2 in STOI, eSTOI, PESQ, and DNSMOS, but improves USES2 by about 0.8 dB in SI-SDR, 4 percent in WER, and 0.17 in UT-MOS. In 8-channels cases, Refined USES2 improves USES2 by 0.5 dB in SI-SDR, 6.3 percent in WER, and about 0.1 in UTMOSv2, without degradation in any metrics.
% This shows ArrayDPS-Refine's effectiveness and robustness when refining a competitive SOTA model.

One interesting observation is that $\xi$ can be a good knob to balance perceptual quality and intelligibility. When $\xi$ increase from $0.4$ to $1.2$, we can in general observe an improvement in intelligibility-based metrics like STOI, eSTOI, and WER, while seeing a degradation in perceptual-based metrics like PESQ, DNSMOS and UTMOSv2.
% For $\xi$ to $1.2$, we sometimes see that all metrics tend to degrade. We believe this is because a larger $\xi$ would bring too much noise to the output.

\vspace{-10pt}
\section{Conclusion}
\vspace{-3pt}
We propose ArrayDPS-Refine, a training-free, generative, and array-agnostic method to improve any discriminative model for multi-channel speech enhancement. We use discriminative model's output to estimate the noise spatial covariance matrix, and then apply ArrayDPS using a pre-trained speech diffusion model. ArrayDPS-Refine shows consistent improvement on multiple discriminative models, including SOTA models, in both perceptual, intelligibility and WER metrics.
\label{sec:refs}

% List and number all bibliographical references at the end of the
% paper. The references can be numbered in alphabetic order or in
% order of appearance in the document. When referring to them in
% the text, type the corresponding reference number in square
% brackets as shown at the end of this sentence \cite{C2}. An
% additional final page (the fifth page, in most cases) is
% allowed, but must contain only references to the prior
% literature.

% Please follow the IEEE Citation Guidelines, \url{https://ieee-dataport.org/sites/default/files/analysis/27/IEEE\%20Citation\%20Guidelines.pdf} for formatting of references.

\bibliographystyle{IEEEbib}
% \ninept
\footnotesize
\bibliography{strings,refs}

\end{document}